\documentstyle[prl,floats,aps,twocolumn,epsf]{revtex}

\tighten

\begin{document}
\draft

\twocolumn[\hsize\textwidth\columnwidth\hsize\csname
@twocolumnfalse\endcsname
\title{Gravitational self-force on a particle orbiting a Kerr black hole}
\author{Leor Barack$^1$ \and Amos Ori$^2$}
\address{
$^1$Albert-Einstein-Institut, Max-Planck-Institut f{\"u}r Gravitationsphysik,
Am M\"uhlenberg 1, D-14476 Golm, Germany\\
$^2$Department of Physics,
Technion---Israel Institute of Technology, Haifa, 32000, Israel}
\date{\today}
\maketitle
\begin{abstract}

We present a practical method for calculating the gravitational self-force,
as well as the electromagnetic and scalar self forces, for a particle in a
generic orbit around a Kerr black hole. In particular, we provide the values
of all the regularization parameters needed for implementing the
(previously introduced) {\it mode-sum regularization} method.
We also address the gauge-regularization problem, as
well as a few other issues involved in the calculation of
gravitational radiation-reaction in Kerr spacetime.

\end{abstract}
\vspace{2ex}
]


The gravitational two-body problem is a long-standing open problem in General
Relativity. In Newtonian theory the evolution of a binary system is easily
analyzed, yielding periodic orbits with conserved energy and angular momentum.
In General Relativity, in contrast, the orbits in a binary system are never
periodic, and the orbital energy and angular momentum decrease due to emission
of gravitational radiation. Typically, the two masses will gradually inspiral
towards each other, until they merge. An interesting branch of the problem
refers to the case where one of the components is a compact object much lighter
than the other. In this situation, the lighter object may be treated as an
infinitesimally small ``particle'' moving on the fixed gravitational-field
background of the heavier object. The particle then experiences a {\em self
force} (SF) due to interaction with its own gravitational field (which is
distorted by the field of the massive object). The issue, then, becomes that of
determining the gravitational SF acting on the particle. The electromagnetic
(EM), flat-space analogue of this problem was explored many decades ago by
Lorentz \cite{Lorentz} and Dirac \cite{Dirac}, who analyzed the force acting on
an accelerating point charge due to interaction with its own EM field. To
analyze the evolution of a strong-gravity, large mass-ratio binary system, the
classic works by Lorentz and Dirac need to be generalized in two important
respects: From the EM to the gravitational SF, and from flat to curved
spacetime.

DeWitt and Brehme \cite{DB} were first to extend the SF concept to
curved spacetime, in the EM case. More recently, a theoretical framework has
been established \cite{MSTQW} for the {\em gravitational} SF in curved
spacetime (cf.\ \cite{alternative}). An essential feature of both the EM and
gravitational SFs in curved spacetime (unlike the analogous flat-space case)
is that they depend, at any given point along the particle's worldline, upon
the entire past history of the particle. This may be attributed to the
scattering of the (EM or gravitational) waves off the spacetime curvature.
So far, this non-locality rendered actual calculations of the SF in curved
spacetime most challenging.

The quest for a practical SF calculation method has considerably intensified
in recent years, with the deployment of a new generation of
gravitational-wave detectors. One of the main sources for the planned
space-based gravitational-wave detector LISA \cite{LISA} would be the
inspiral of compact objects [white dwarfs, neutron stars, and stellar black
holes (BHs)] into super-massive BHs, with a typical mass ratio of $10^{3}$-$%
10^{7}$. Such super-massive BHs appear to reside in the centers of many
galaxies, including our own \cite{MilkeyWay}. In a typical scenario, the compact
object spends the last year of inspiral deep inside the highly relativistic
region near the BH's event horizon, emitting some $10^{5}$ cycles of
gravitational waves \cite{FT}, which carry a detailed information of the
massive BH's strong-field geometry \cite{Ryan}. Knowledge of the self
force---which, in turn, determines the orbital evolution---is necessary for
predicting the emitted waveform. Such a prediction is also important for
the design of templates which should significantly improve the chance of
detection in the situation of a small signal-to-noise ratio, typical to the
first generation of detectors.

In several, relatively simple situations the orbit's evolution may be
inferred using standard energy-momentum balance methods \cite{balance}.
Namely, one calculates the flux of energy and angular momentum carried away
by the gravitational waves, and equates this to (minus) the rate of change
in the corresponding entities associated with the orbiting particle itself.
This simple method yields the orbital evolution
for all orbits around a non-spinning (Schwarzschild type) BH, and also for
strictly equatorial or circular orbits around a rotating (Kerr type) BH.
However, in the astrophysically relevant situation one needs to analyze the
evolution of {\em generic} orbits around a Kerr BH. In this generic case,
energy-momentum balance methods are insufficient, because the energy and
(azimuthal) angular-momentum no longer uniquely determine the orbit.
One therefore must calculate the local SF acting on the
particle. The goal of this paper is to provide a practical method for
calculating this SF, for generic orbits in Kerr spacetime.


Based on the basic formulation of Refs.\ \cite{MSTQW}, we previously devised
the {\em mode-sum method}, as a practical technique for SF calculations \cite
{MSRS,MSRS-implement}. In this method, one first calculates the (finite)
contribution to the SF due to individual multipole modes of the particle's
field (this is done by integrating numerically the mode-decomposed
linearized field equations). Then, a certain regularization procedure is
applied to the mode sum [see Eq.\ (\ref{eq10}) below], which involves
certain ``regularization parameters'' (RP). The latter depend upon the
background spacetime and the specific orbit under consideration. The RP
values were previously derived for equatorial orbits around a
Schwarzschild BH \cite{Letter,details1,details2,MNS}. Here we extend
the analysis and give the RP values for
an arbitrary geodesic orbit around a Kerr BH.


Throughout this Letter we use units where $c=G=1$, metric signature ${-}{+}{+%
}{+}$, and Boyer--Lindquist coordinates $t,r,\theta ,\varphi $. For
completeness, we shall address here not only the gravitational SF, but also
the analogous EM and scalar phenomena.
We shall thus consider the following
setup: A particle of mass $\mu $, electric charge $e$, and/or scalar charge $%
q$ moves along an arbitrary free-fall orbit (i.e., no external forces) $%
z^{\mu }(\tau )$ around a Kerr BH with mass $M\gg \mu ,|q|,|e|$ and
arbitrary spin $aM$. We shall consider the SF acting on the particle
(to the leading order $\propto \mu ^{2},e^{2},q^{2}$) at an
arbitrary point $z_{0}^{\mu }=(t_{0},r_{0},\theta _{0},\varphi _{0})$ along
its trajectory. 
The particle's equation of motion, including the SF effect, reads
\begin{equation}
\mu u_{\lambda ;\beta }u^{\beta }=F_{\lambda }^{{\rm self}},  \label{eq2}
\end{equation}
where $u^{\beta}\equiv dz^{\beta}/d\tau $ is the four-velocity, a semicolon
denotes covariant differentiation with respect to the background (Kerr)
geometry, and $F_{\lambda }^{{\rm self}}$ is the SF.

Let $x$ denote a spacetime point in the neighborhood of $z_{0}$. We denote
the particle's actual perturbation field (the ``full field'') by $h_{\alpha
\beta }(x)$, ${\cal A}_{\alpha }(x)$, or $\Phi (x)$, representing,
correspondingly, the metric perturbation ($\propto \mu $), EM vector
potential ($\propto e$), and scalar field ($\propto q$). These fields are,
in principle, solutions to the linearized Einstein/Maxwell/Klein-Gordon
equations, with the particle serving as a source term. To these fields there
correspond ``full-force'' fields, given by \cite{details1,details2}
\begin{equation} \label{eq5}
F_{\mu }^{{\rm full}}(x)=\left\{
\begin{array}{ll}
q\Phi _{,\mu }, & \text{(scalar)}, \\
e\kappa_{\mu }{}^{\beta \gamma }{\cal A}_{\beta ;\gamma }, & \text{(EM)}, \\
\mu\kappa_{\mu }{}^{\beta \gamma \delta }{h}_{\beta \gamma ;\delta }, &
\text{(grav.)},
\end{array}
\right.
\end{equation}
where
the tensors $\kappa$ are
\begin{eqnarray}  \label{eq7}
\kappa^{\alpha \beta \gamma }(x) &\equiv &g^{\alpha \gamma }\hat{u}^{\beta
}-g^{\alpha \beta }\hat{u}^{\gamma },  \nonumber  \\
\kappa^{\alpha \beta \gamma \delta }(x) &\equiv&
-g^{\alpha\beta}\hat{u}^{\gamma}\hat{u}^{\delta}
+\frac{1}{2}g^{\alpha\delta}\hat u^{\beta}\hat u^{\gamma}
-\frac{1}{2}\hat u^{\alpha}\hat u^{\beta}\hat u^{\gamma}\hat u^{\delta}.
\end{eqnarray}
Here $g^{\alpha \beta }$ is the background metric at $x$ and $\hat{u}%
^{\alpha }$ is the four-velocity parallelly propagated from the worldline to
$x$ along a short geodesic normal to the worldline.

In the mode-sum method, the SF is then calculated through \cite{MSRS}
\begin{equation} \label{eq10}
F_{\mu }^{{\rm self}}=\sum_{l=0}^{\infty }\left[ (F_{\mu l}^{{\rm full}%
})^{\pm }-A_{\mu }^{\pm }L-B_{\mu }-C_{\mu }/L\right] -D_{\mu }.
\end{equation}
Here $L\equiv l+1/2$, and $(F_{\mu l}^{{\rm full}})^{\pm }$ is the $l$
multipole mode of $F_{\mu }^{{\rm full}}(x)$, summed over all azimuthal
numbers $m$ (for a given $l$) and evaluated at ($t_{0},r\to r_{0}^{\pm
},\theta _{0},\varphi _{0}$). The full modes $F_{\mu l}^{{\rm full}}$ are
generally discontinues across $r=r_{0}$, hence the label $\pm $ \cite{MSRS}.
The RP, i.e., the quantities $A_{\mu }^{\pm }$, $B_{\mu }$, $C_{\mu }$, $%
D_{\mu }$, are to be derived analytically (by analyzing the local singular
behavior of the perturbation field near the particle). The first three of
these parameters capture the asymptotic form of the full-force modes at
large $l$, making the sum in Eq.\ (\ref{eq10}) convergent. The parameter $%
D_{\mu }$ is a certain residual quantity that arises in the summation over $%
l $---for a more precise definition of the RP see \cite{MSRS,Letter}.

In deriving the RP in Kerr spacetime, we basically followed the method used
in the Schwarzschild case \cite{Letter}, as described in much detail in
Refs.\ \cite{details1,details2}. Namely, we considered the direct part of
the full-force field (i.e., the part associated with waves directly
propagating along the light cone), from which all RP are obtained by
applying the multipole decomposition and appropriately taking the limit $%
x\to z_{0}$. The main technical challenge in extending our analysis to the
Kerr case lies in the fact that one can no longer exploit spherical symmetry
by choosing a coordinate system in which the orbit is confined to the
equatorial plain. Rather, one must carry out the calculation at an arbitrary
value of $\theta _{0}$. This turns out to render the entire analysis
significantly more complicated. For lack of space, we
skip the derivation of these RP values---the detailed derivation will be
given elsewhere \cite{KerrDetailed}. In what follows we merely present the
results, i.e., the values of all RP for an arbitrary geodesic in Kerr
spacetime.

To write down the RP in a convenient unified form, we introduce a
``generalized'' charge $q_{s}$, where $q_{s}=q$, $e$ or $\mu $, for the
scalar ($s=0$), EM ($s=1$), and gravitational ($s=2$) cases, respectively.
All RP are then independent of $s$, apart from an overall
factor $(-1)^{s}q_{s}^{2}$. For $A_{\mu }$, $C_{\mu }$, and $D_{\mu }$ we
find
\label{RP}\label{CD}
\begin{equation}
C_{\mu }=D_{\mu }=0,
\end{equation}
and 
\begin{eqnarray} \label{A}
A_{r}^{\pm } &=&\mp (-1)^{s}\frac{q_{s}^{2}}{V}\left( \frac{\sin ^{2}\theta
_{0}\,g_{rr}}{g_{\theta \theta }g_{\varphi \varphi }}\right) ^{1/2}\left(
V+u_{r}^{2}/g_{rr}\right) ^{1/2},  \nonumber   \\
A_{t}^{\pm } &=&-(u^{r}/u^{t})A_{r}^{\pm },\quad \quad A_{\theta }^{\pm
}=A_{\varphi }^{\pm }=0,
\end{eqnarray}
where hereafter $g_{\alpha \beta }\equiv g_{\alpha \beta }(z_{0})$ and
$V\equiv 1+u_{\theta }^{2}/g_{\theta \theta }
+u_{\varphi }^{2}/g_{\varphi \varphi }$.
The expression for $B_{\mu }$ takes a slightly more complicated form,
\begin{equation}
B_{\mu }=(-1)^{s}q_{s}^{2}(2\pi )^{-1}P_{\mu abcd}I^{abcd}.  \label{B}
\end{equation}
Hereafter, roman indices ($a,b,c,...$) run over the two Boyer--Lindquist
angular coordinates $\theta ,\varphi $. The coefficients $P_{\mu abcd}$ are
given by 
\begin{eqnarray}\label{eq60}
P_{\mu abcd} &=&(3P_{\mu a}P_{be}-P_{\mu e}P_{ab})C_{cd}^{e}  \nonumber
 \\
&&+\frac{1}{2}\left[ 3P_{\mu d}P_{abc}-(2P_{\mu ab}+P_{ab\mu })P_{cd}\right]
,
\end{eqnarray}
where
\begin{eqnarray} \label{eq70}
P_{\alpha \beta } &\equiv &g_{\alpha \beta }+u_{\alpha }u_{\beta },
\nonumber   \\
P_{\alpha \beta \gamma } &\equiv &\left( u_{\lambda }u_{\gamma }\Gamma
_{\alpha \beta }^{\lambda }+g_{\alpha \beta ,\gamma }/2\right)
\end{eqnarray}
($\Gamma _{\alpha \beta }^{\lambda }$ denoting the connection coefficients
at $z_{0}$), and
\begin{equation}  \label{C}
C_{\varphi \varphi }^{\theta }=\frac{1}{2}\sin \theta _{0}\cos \theta
_{0},\quad C_{\theta \varphi }^{\varphi }=C_{\varphi \theta }^{\varphi }=-%
\frac{1}{2}\cot \theta _{0},
\end{equation}
with all other coefficients $C_{cd}^{e}$ vanishing.
Finally, the quantities $I^{abcd}$ are
\[
I^{abcd}=(\sin \theta _{0})^{-N}\int_{0}^{2\pi }G(\gamma )^{-5/2}(\sin
\gamma )^{N}(\cos \gamma )^{4-N}\,d\gamma ,
\]
where $N\equiv N(abcd)$ is the number of times the index $\varphi $ occurs
in the combination $abcd$ (namely,
$
N=\delta _{\varphi }^{a}+\delta _{\varphi }^{b}+\delta _{\varphi
}^{c}+\delta _{\varphi }^{d}
$)
and 
\begin{equation}
G(\gamma )=P_{\tilde{\varphi}\tilde{\varphi}}\sin ^{2}\gamma +2P_{\theta
\tilde{\varphi}}\sin \gamma \cos \gamma +P_{\theta \theta }\cos ^{2}\gamma ,
\label{G}
\end{equation}
where
\begin{equation}
P_{\tilde{\varphi}\tilde{\varphi}}\equiv P_{\varphi \varphi }/\sin
^{2}\theta _{0},\quad \quad P_{\theta \tilde{\varphi}}\equiv P_{\theta
\varphi }/\sin \theta _{0}.  \label{tilde}
\end{equation}
The integrals $I^{abcd}$ can be expressed in terms of complete Elliptic
integrals. These relations are given explicitly in Ref.\ \cite{gr-qc}.

In the special case of an equatorial orbit in Schwarzschild spacetime,
the above RP values reduce to those obtained previously using various
methods \cite {MSRS,Letter}.

In the rest of this paper we discuss several important issues that arise
when implementing the mode-sum method in Kerr spacetime.

First, to avoid confusion, we emphasize that the quantities $F_{\mu l}^{{\rm %
full}}$ appearing in Eq.\ (\ref{eq10}) refer (in all three cases $s=0,1,2$)
to the decomposition of the full-force field components $F_{\mu }^{{\rm full}%
}$ in the standard {\em scalar spherical harmonics}. Practically, in the
Kerr case one separates the field equations in the frequency domain, using
(spin weighted) {\em spheroidal harmonics}. Then the full-force field, too,
is obtained in terms of the spheroidal harmonics. Thus, in order to
implement the mode-sum scheme (at least in its present formulation), one
needs to decompose the contributions to the full force from the various
spheroidal-harmonic modes, into scalar spherical harmonics:
For a given $l$ (and given azimuthal number $m$ and temporal frequency
$\omega $) one collects the contributions to the spherical-harmonic
mode $l,m$ coming from the full-force fields associated with the
spheroidal-harmonic modes $l^{\prime },m,\omega $. Then one sums over
$l^{\prime }$ (as well as $m,\omega $) to obtain $F_{\mu l}^{{\rm full}}$.
The decomposition of the spheroidal harmonics in spherical harmonics
is described in Ref.\ \cite{Hughes}.

Our second remark concerns the construction of the full-force field in the
EM and gravitational cases. This field involves an extension of $u^{\alpha }$
off the worldline, cf.\ Eq.\ (\ref{eq7}). Here we adopted the most
natural extension $\hat{u}^{\alpha }$, namely, the one obtained by parallelly
propagating $u^{\alpha }$ from the worldline to $x$. As discussed in \cite
{details2}, it is possible to use any other (sufficiently regular) extension
of $u^{\alpha }$. Note, however, that changing the extension (which
obviously affects the quantities $F_{\mu l}^{{\rm full}}$) will generally
affect the value of the parameter $B_{\mu }$ (though it turns out that the
other RP are unaffected). Nevertheless, if two extensions share the same
values of angular derivatives $u_{,b}^{\alpha }$ at the worldline, they
admit the same $B_{\mu }$ \cite{details2}.

The parallelly-propagated extension is an elegant one, and, furthermore, it
allows the simplest derivation of $B_{\mu }$. However, this extension is rather
inconvenient for calculating the full-force quantities $F_{\mu l}^{{\rm full}%
}$ (the Legendre decomposition requires the extension of $u^{\alpha }$ far
away from the worldline, on the entire two-sphere $r,t$=const; It is hard
to calculate $\hat{u}^{\alpha }$ explicitly off the worldline, and it is even
harder to Legendre-decompose it. Furthermore, there is no guarantee that $\hat{u}%
^{\alpha }$ would be globally well defined). To overcome this difficulty, we
propose the following strategy: Compose any global extension $\tilde{u}%
^{\alpha }$ off the worldline, with the only demand that $\tilde{u}%
_{,b}^{\alpha }=\hat{u}_{,b}^{\alpha }$ (and, obviously, $\tilde{u}^{\alpha
}=\hat{u}^{\alpha }$) at the worldline itself. It is easy to construct such simple
global extensions. Then simply use this $\tilde{u}^{\alpha }$ instead of $%
\hat{u}^{\alpha }$ in Eq.\ (\ref{eq7}). The above values of all RP are
unaffected by this change of extension.

There is another technical issue that arises in the EM and gravitational
cases: The only known scheme for separating the EM and (linearized)
gravitational field equations in Kerr spacetime is the Teukolsky formalism
\cite{Teukolsky}. By solving the Teukolsky equation, one obtains certain
components of the Maxwell or Weyl tensors, respectively. However, as is
obvious from Eq.\ (\ref{eq5}), the calculation of $F_{\mu }^{{\rm full}}$
(and hence of $F_{\mu l}^{{\rm full}}$ and the SF) requires the knowledge of
the basic perturbation fields, $h_{\alpha \beta }(x)$ or ${\cal A}_{\alpha }(x)$.
A method for reconstructing $h_{\alpha \beta }$ and ${\cal A}_{\alpha }$ for
a point particle, out of the Teukolsky variables, based on the Chrzanowski-Wald
formalism \cite{Chrzanowski,Wald}, was recently provided by Ori \cite{Ori}
(see also \cite{Lousto&Whiting} for the Schwarzschild case).
More recently, this construction was extended to the non-radiative modes
$l=0,1$, as well as to all higher-$l$ modes with $\omega =0$ \cite{Ori2}.

Finally, we address the crucial issue of the {\em gauge-dependence} of the
gravitational SF, and the related problem of gauge regularization in Kerr
spacetime. The gravitational SF (unlike the scalar and EM forces) turns out
to be gauge-dependent \cite{gauge}. The basic formalism \cite{MSTQW}
is given in terms of the harmonic gauge. The transformation of the SF to any
other gauge is given in Ref. \cite{gauge}. This transformation, however, is
restricted to {\em regular gauges}, i.e., gauges related to the harmonic
gauge via a displacement vector $\xi^{\mu}$ that is continuous on the particle's
worldline. (If the gauge is irregular, then the SF generally becomes ill-defined.) For
any regular gauge, the SF may, in principle, be calculated through Eq.\ (\ref
{eq10})---provided that one uses the quantities $F_{\mu l}^{{\rm full}}$
associated with this same gauge. The RP are gauge-invariant in this context
\cite{gauge}.

Although the momentary SF is gauge-dependent, the long term
radiation-reaction evolution of the orbit, as expressed, e.g., by the drift of
the constants of motion, is gauge-invariant. The radiation-reaction
evolution can therefore be calculated in any regular gauge.

Chrzanowski's construction yields the metric perturbations in the so-called
{\em radiation gauge} \cite{Chrzanowski}. Unfortunately, this gauge is not
regular, hence the SF does not have a well-defined value in the radiation
gauge. To overcome this difficulty, we previously proposed\cite{gauge} to
calculate the SF in an {\em intermediate gauge}, namely, a regular gauge
which (unlike the harmonic gauge) can be obtained from the radiation gauge
via a simple, explicit gauge transformation. A convenient such intermediate
gauge was recently constructed explicitly in Ref.\ \cite{Ori2}.

When implementing the mode-sum scheme (\ref{eq10}) to the SF in the
intermediate gauge, one should, in principle, substitute the quantities
$F_{\mu l}^{{\rm full(INT)}}$ associated with this same gauge. Alternatively,
one may still use the radiation-gauge quantities $F_{\mu l}^{{\rm full(RAD)}}$
in Eq.\ (\ref{eq10}), but in this case one must subtract an extra term,
which we denote $\delta D_{\mu }$, to compensate for the differences $\delta
F_{\mu l}^{{\rm full}}$ in the values of $F_{\mu l}^{{\rm full}}$ in the two
gauges. (The derivation of $\delta F_{\mu l}^{{\rm full}}$ from $\xi^{\mu }$
is outlined at the end of Ref. \cite{gauge}; $\delta D_{\mu }$ is the sum of
$\delta F_{\mu l}^{{\rm full}}$ over $l$.) Thus, when applied to the SF in
the intermediate gauge, Eq.\ (\ref{eq10}) may be recast as
\[
F_{\mu }^{{\rm self(INT)}}=\sum_{l=0}^{\infty }\left[ (F_{\mu l}^{{\rm %
full(RAD)}})^{\pm }-A_{\mu }^{\pm }L-B_{\mu }\right] -\delta D_{\mu }
\]
(recall $C_{\mu }=D_{\mu }=0$). The RP $A_{\mu }^{\pm },B_{\mu }$ are those
given in Eqs.\ (\ref{A}) and (\ref{B}) (as mentioned above, the RP values are
gauge-invariant). The term $\delta D_{\mu }$ was recently calculated for a
generic orbit in Kerr, and its expression
(whose structure and complexity
resemble those of $B_{\mu }$ above) will be given elsewhere \cite{Ori2}.
This provides the solution to the gauge-regularization problem. The
quantity $F_{\mu }^{{\rm self(INT)}}$ can now be used to analyze the
(gauge-independent) long-term radiation-reaction evolution, for any orbit in
Kerr spacetime.

L.B.\ was supported by a Marie Curie Fellowship of the European Community
program IHP-MCIF-99-1 under contract number HPMF-CT-2000-00851. A.O. was
supported by The Israel Science Foundation (grant no.\ 74/02-11.1).

\end{document}